\documentstyle[prl,aps,twocolumn,psfig]{revtex}

\begin{document} \title{Control of Spin Dynamics of Excitons in Nanodots for
Quantum Operations}
\twocolumn[\hsize\textwidth\columnwidth\hsize\csname@twocolumnfalse\endcsname
\author{Pochung Chen, C. Piermarocchi, and L. J. Sham}
\address{Department of Physics, University of California San Diego, La Jolla CA
92093-0319.}

\date{\today} \maketitle

\draft


\begin{abstract} 

This work presents a step furthering a new perspective of proactive
control of the spin-exciton dynamics in the quantum limit. Laser
manipulation of spin-polarized optical excitations in a semiconductor
nanodot is used to control the spin dynamics of two interacting
excitons.  Shaping of femtosecond laser pulses keeps the quantum
operation within the decoherence time.  Computation of the fidelity of
the operations and application to the complete solution of a basic
quantum computing algorithm demonstrate in theory the feasibility of
quantum control.  

\end{abstract} \pacs{} \date{\today} ]

Experimental laser control of spin states of excitons has been
demonstrated in ensembles of semiconductor quantum dots (QDs)
\cite{awsch}. In a single dot, ultrafast control of
spin-excitons\cite{bonadeo}, and entanglement of the electron-hole
complex \cite{chen} have been reported. Here, we present a study of
the design of laser pulses for the ultrafast control of the spin
dynamics of individual excitons in a QD. This is a special case of
designing Hamiltonians to bring a system from one state to another
\cite{warren}. We hope that the theory of control of exciton spin
dynamics may help rapid realization of basic quantum operations in
nanodots. An exciting application would be the further development of
quantum computation \cite{chuang} using excitons
\cite{imamoglu00,wang,troiani,biolatti,pchen} or intersubband
transitions \cite{barenco,sanders00,sherwin}.

In this paper, the control of exciton spin dynamics consists in
designing laser pulses to create as a function of time desired
multiexciton states of a subsystem of dots.  In the ultrafast control
of the spin dynamics of excitons, there is the possibility of
unintended dynamics due to the presence of close resonances. Sharp
resonant pulses can minimize such unintended dynamics, but at the cost
of a long operation time. The latter leads to the spontaneous
recombination of the excitons and to their dephasing, resulting in
uncontrolled deterioration of the amplitude and phase of the
coefficients of the desired linear combination of quantum states. We
give a solution to these two contradictory requirements by suggesting
extending the well-known laser pulse-shaping technique \cite{weiner}
 to the delicate control in the quantum limit.  Design of excitations
has been applied to quantum computation by NMR \cite{cummins} and by
electron spin dynamics or intersubband transitions in QDs
\cite{berman,burkard99,tian,sanders99}.  We rely on the theory of
DiVincenzo \cite{divince} stating that quantum gates operating on just
two qubits at a time are sufficient to construct a general quantum
circuit in a system of many dots. As an application to a prototype
quantum computation which illustrates the issues raised above, we
present the results of numerical simulations of the two exciton
dynamics in a sequence of operations to solve the Deutsch-Jozsa (DJ)
problem \cite{deutsch,chuang}.  We shall concentrate on the control of
the lowest four states formed by two excitons with opposite spins in a
single dot. The spin-down and up states can be excited by the left
($\sigma-$) and right-handed ($\sigma+$) circularly polarized light,
respectively. The four basis states are, in order: the lowest
biexciton state $|+-\rangle$, the $\sigma+$ exciton state $|+\rangle$,
the $\sigma-$ exciton state $|-\rangle$, and the ground state
$|0\rangle$. We use the interaction representation for the
states. We allow for the presence of other multiexciton states in the
dot and the connection of the dot to the environment.

\begin{figure}
\centerline{\psfig{file=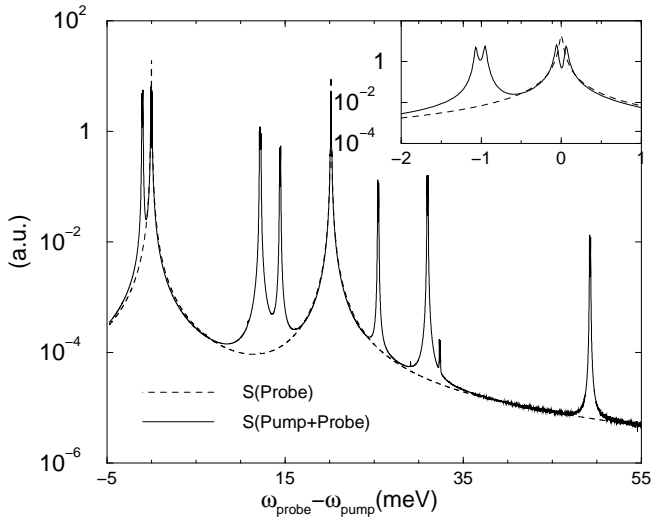,width=8.6 cm}} \caption{Coherent
nonlinear spectra of a QD with pump and probe in opposite circular
polarization. The pump is resonant with the lowest excitonic state.
The solid line is for the pump beam with intensity corresponding to a
Rabi energy of 50~$\mu$eV and the dashed line for zero pump
intensity. The inset is a zoom on the first two
levels. }  \label{cwspectra} 
\end{figure}

For precise control of coherent exciton dynamics, the probabilistic
photo-excitation process of an exciton needs to be replaced by a
resonant Rabi rotation which can create a linear combination of two
states at a definite time. The primary condition for carrying out a
Rabi rotation is sufficient light intensity without generating more
excitons. This can be ascertained by analyzing the coherent nonlinear
optical spectra. The starting point of our calculation is the energies
and wave functions of the multi-exciton states in a dot.  For the
current purpose, it is sufficient to obtain these states from two
confined levels of electrons and holes each in a parallelepiped QD
\cite{barencodup95}. The size of the dot,
35~nm$\times$40~nm$\times$5~nm, is typical of interface fluctuation
quantum dots \cite{bonadeoprl98}.  Only Coulomb interaction between
the carriers which conserves their conduction or valence band indices
is taken into account exactly.  The resulting states are used to
compute the nonlinear spectra in a coherent pump and probe experiment
setup as in \cite{bonadeoprl98}.  Each cw light beam has a narrow
frequency peak and a definite circular polarization.  The two beams
have a fixed phase difference. The nonlinear spectrum is the
absorption of the probe beam as a function of its frequency for a
given intensity and frequency of the pump beam. An example of the
calculated spectrum is shown in Fig.~\ref{cwspectra} for the
cross-polarized circular configuration.  The peaks come from
transitions between the multi-exciton states. The key to our quantum
operations is the lowest two peaks, corresponding to the
exciton-biexciton and the ground state-exciton transitions,
respectively (magnified in the inset). The separation of 1~meV between
the two peaks corresponds to the binding energy of $|+-\rangle$. An
exciton lifetime of 40 ps is taken from the experiment
\cite{bonadeoprl98}. The doublet structure in each peak is the Rabi
splitting due to the presence of the strong pump field.  The
observation in the nonlinear spectrum of these doublets would indicate
the feasibility of Rabi rotations in real-time operations.  The
life-time broadening and the line positions yield the time limit on
the pulses to be used in the exciton operations. The observation of
Rabi rotations in QDs has not yet been reported. Evidence of Rabi
flopping for excitons in semiconductor quantum wells
\cite{schulzgen99} is encouraging.

\begin{figure} 
\centerline{\psfig{file=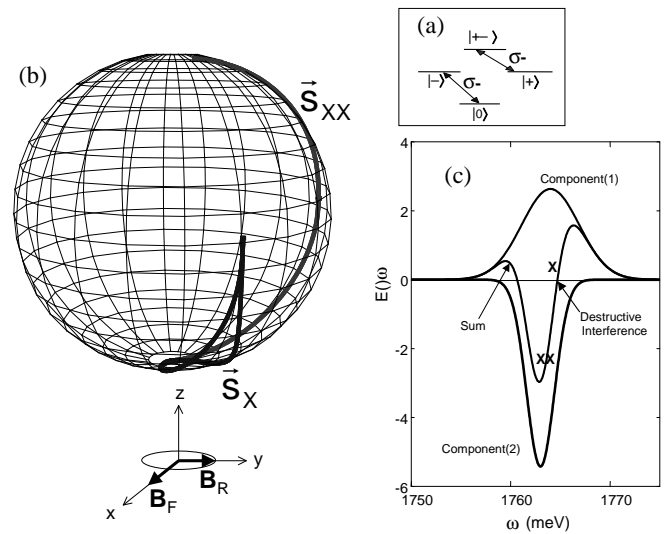,width= 8.6 cm,angle=
90}}\vspace{0.5 cm} 
\caption{Dynamics of spectral shaping.  (a) Energy
schematics for the transitions generated by a $\sigma-$ pulse.  (b)
Evolution of the exciton $\vec{s}_X$ and biexciton $\vec{s}_{XX}$
pseudospins on the Bloch sphere under a shaped pulse. (c) Fourier
transform of the shaped pulse and its components.} 
\label{optim}
\end{figure}

The first quantum operation considered here is the conditional
dynamics where the $\sigma-$ exciton will be given a $\pi$ rotation
only if there is already a $\sigma+$ exciton present.  This controlled
Rabi rotation (C-ROT) is related to the controlled-not (C-NOT) or
exclusive-or logic operation between two qubits \cite{chuang}. A C-NOT
operation inverts the second qubit (the target) depending on the state
of the first qubit (the control).  The C-ROT operation is accomplished
simply by using a $\sigma-$ circularly polarized laser pulse to effect
a single resonant Rabi $\pi$-rotation \cite{lloyd,barenco,troiani}
which exchanges the exciton state $ |+\rangle $ with the biexciton
state $-|+-\rangle$ [see Fig.~\ref{optim}(a)].  Ideally, the same
pulse does not cause the transition between $|0\rangle$ and
$|-\rangle$ because the interaction between two excitons cause this
transition to be off-resonance by an amount equal to the binding
energy of the biexciton. The second operation is a single qubit
operation: it acts on an exciton regardless of the state of the other
exciton. It requires a resonant Rabi rotation for each of the many
transitions involving the exciton. In the 2-exciton subspace, an
operation on, say, the $\sigma-$ exciton requires two frequencies for
the two transitions between the ground state $|0\rangle$ and $
|-\rangle$ and between $ |+\rangle$ and $|+-\rangle$, i.e., a
two-color laser pulse is needed.  The idea of a resonant Rabi rotation
for C-NOT was proposed by Lloyd \cite{lloyd} and by Barenco et al,
\cite{barenco}.  To complete the C-NOT, we could add to a C-ROT
additional operations. However, we have found no need in quantum
algorithms to take this time consuming extra step. Proper rotations
represent the natural choice for the physical design of algorithms
using Rabi flopping.  We will show below how C-NOT and Hadamard gates
in the standard implementation of the DJ algorithm \cite{chuang} can
be replaced by C-ROT and $\pi/2$ operations. Using proper rotations,
we have also designed the optical control for quantum Fourier
transform, which is key to several algorithms including factorization,
and carried out the simulation\cite{pc01}.

We have simulated these two basic operations and calculated the gate
fidelity defined in Ref. \cite{poyatos}, whose equally weighted
average over all the possible initial states gives
$\frac{1}{10}\sum_{i}|I_{ii}|^2+\frac{1}{20}\sum_{i\ne j} (I^*_{ii}
I_{jj}+I^*_{ij} I_{ij})$. $I_{ij}=\langle i|\bar{U}^\dagger
U|j\rangle$, where $\bar{U}$ and $U$ are the calculated and the ideal
quantum transformations respectively.  We have included in addition to
the four states in the computational basis the two copolarized
biexciton states $|++\rangle,|--\rangle$ in order to check their
contribution to the unintended dynamics. The other states are more
distant in energy and can be neglected.  The dephasing is taken into
account phenomenologically as a stochastic process using the Monte
Carlo wavefunction method \cite{jumps}. A 40~ps exciton lifetime is
taken from \cite{bonadeoprl98}. If we use a spectrally sharp pulse to
perform C-ROT, without dephasing the unintended perturbation is
negligible but the operation time is of the same order as the
dephasing time. On the other hand, a sufficiently short pulse (about
2~ps) to limit dephasing destroys the fidelity by unintended dynamics.
We propose a remedy by pulse shaping.  Let us first compare the
results in fidelity of the operations before presenting the details of
shaping. The fidelity decreases with increasing Rabi frequency. The
calculated results used 2~meV Rabi energy which is larger than the
separation between the exciton and biexciton resonances of 1~meV. The
fidelity for C-ROT is 0.535, 0.966, and 0.903 for, in order, unshaped
short pulse (without dephasing), shaped pulse without and with
dephasing. For single qubit $\pi/2$-rotation, it is 0.678, 0.995 and
0.981 respectively. The fidelity is nicely restored by pulse shaping.
The leakage to the unwanted states$|++\rangle,|--\rangle$ is less than
1\%.

In a C-ROT, instead of a spectrally narrow $\pi$ pulse resonant with
the exciton-biexciton transition, we use a combination of two
phase-locked pulses of the $\sigma-$ polarized field, \begin{equation}
{\cal E}(t)={\cal E}_0(e^{-(t/t_1)^2-i\epsilon_-
t}+e^{-(t/t_2)^2-i(\epsilon_- - \Delta_{+-})t + i \pi})~.
\label{pulse} \end{equation} For convenience, we use equal
amplitudes. The energy of each state is denoted by $\epsilon$ with an
index for that state.  The biexciton binding energy is $\Delta_{+-} =
\epsilon_{+} +\epsilon_{-} - \epsilon_{+-}$. The dynamics of
the four exciton levels is shown in Fig.~\ref{optim}, visualized as
pseudospin (Bloch vector) for the $\sigma-$ exciton in the absence or
presence of the $\sigma+$ exciton (see insert (a)). In the absence of
$\sigma+$, the $\sigma-$ exciton pseudospin evolves in an effective magnetic
field ${\bf B}_{X}(t)={\bf B}_F(t,t_1)+{\bf B}_R(t,t_2)$,
 composed of a component ${\bf B}_F$ fixed in the $x-$direction due to
the resonant pulse with the $|0\rangle \leftrightarrow |-\rangle$
transition and another ${\bf B}_R$, due to the off-resonant pulse with
the $|+\rangle \leftrightarrow |+-\rangle$ transition, which rotates
clockwise in the $xy-$plane at the beat frequency $\Delta_{+-}$. In
the presence of the $\sigma+$, the pseudospin is driven by an
effective ${\bf B}_{XX}(t)={\bf B}_R(t,t_1)+{\bf B}_F(t,t_2)$, in a
rotating frame where the field component resonant with the biexciton
resonance is fixed and the exciton field rotates anti-clockwise. The
central idea is to optimize the control parameters ($t_1$, $t_2$) in
such a way that, at the end of the shaped pulse, ${\bf B}_{XX}$
produces a $\pi$ rotation for the biexciton resonance, and ${\bf B}_X$
brings the exciton pseudospin back to the original state. We have
studied numerically the dynamics of the four levels  in the
control manifold generated by the parameters ($t_1$, $t_2$). Even
using Rabi energies larger than the biexcitonic binding energy, it is
possible to find a region in the phase space ($t_1$,$t_2$) fulfilling
the required conditions. This is illustrated in Fig.~\ref{optim}(b),
where the evolution of the pseudospin in the absence and presence of
the $+$ exciton using a pulse shaped as in Eq.~(\ref{pulse}), with
$t_1= 0.56 $~ps and $t_2=1.05$~ps, is represented on the Bloch sphere:
at the end of the shaped pulse we have a $\pi$ rotation for $|+\rangle
\leftrightarrow |+-\rangle$, while $|0\rangle$ remains unaffected. As
inferred from the calculation of the fidelity, this shaping works for
an arbitrary initial state, and the pulse acts as a general rotor.  A
simple intuitive picture is given by the frequency spectrum of the
pulse in Eq.(\ref{pulse}), shown in Fig.~\ref{optim}(c). The two
phase-locked components have opposite phases and the chosen $t_1$ and
$t_2$ control the interference in the electric field giving a zero
amplitude at the exciton resonance and a finite value at the biexciton
resonance. The ultrafast manipulation of the qubits is carried out
using light interference. For the $\pi/2$ single qubit gate we used
$t_1=t_2=0.15$ ps and in-phase components. The full theory for the
application of the shaping to more general operations and limitations
will be provided elsewhere\cite{pc01}.

A cogent illustration of the time limitation of decoherence and the
designs to beat the limit is given by a numerical simulation of a
complete quantum computing algorithm.  We choose the two-qubit DJ
problem \cite{deutsch} both for its simplicity and for its fundamental
nature. It has been implemented by nuclear magnetic resonance (NMR)
\cite{chuang98}. For a single qubit variable $x$, there are four
possible single-qubit functions: $f_1(x)=0$ and $f_2(x)=1$ belong to
the class of constant functions and $f_3(x)=x$ and $f_4(x)=1-x$ the
class of balanced functions.  The DJ problem is to determine whether a
given function is constant or balanced.  Classically, the only way to
find out if the function is constant or balanced is to calculate the
function for both the 0 and 1 input values and then compare the
results. The quantum algorithm designs a unitary transformation which
parallel-processes all possible values of $x$. We use the presence or
absence of $\sigma\pm$ excitons to represent the two qubits.  The
complete solution consists of three steps: (i) {\it Encoding}
transforms the initial ground state dot into an input state:
$|in\rangle=\frac{1}{2}(-|+-\rangle+|+\rangle-|-\rangle+|0\rangle)$ by
the single-qubit $\sigma+$ $-\pi/2$ rotation and $\sigma-$ $\pi/2$
rotation. (ii) The unitary transformation $U_{f_j}|y,x\rangle =
R_{x}(\pi)^{1-2f_j(y)}|y,x\rangle$, associated to the given $f_j$ is
applied.This is a conditional rotation acting on qubit $x$ ($\sigma-$
exciton), conditioned on the effect of $f_j$ on the qubit $y$
($\sigma+$ exciton).  (iii) {\it Decoding} by $\pi/2$ rotations of
both circular polarizations, ending with the ground state if the
function is a constant and the $|+\rangle$ state if the function is
balanced.  The final measurement on the $\sigma+$ exciton can be made
with the cross-polarization probe or with the emitted photon
polarization. 

\begin{figure}
\centerline{\psfig{file=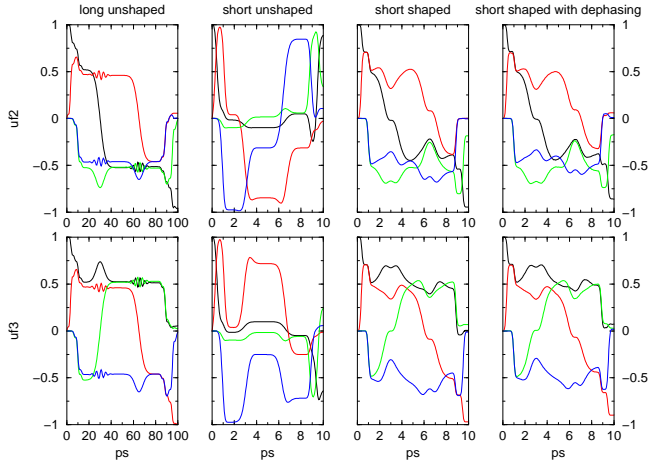,width= 8.6 cm,angle= 270}}
 \caption{Evolution of the qubits during a quantum computation. The
real parts of the coefficients ${\cal R}c_j(t)$ for the first four
levels are plotted.  See text for explanation.}  \label{computing}
\end{figure}

The evolution of the qubits through the DJ algorithm is shown in
Fig.~\ref{computing} as the real parts of the coefficients of the
corresponding states in two rows for a constant ($f_2$) and a balanced
($f_3$) function. The first column depicts the ideal case of sharp
resonance functions without dephasing as described above.  Even with
the maximum packing of the pulses (using overlap at 10\% of the peak
electric field without substantial deterioration of the results), the
first column show the time to complete the quantum computation above
the adopted decoherence time of the order of 40~ps. Simple reduction
of the pulse time to within dephasing time while keeping the pulse
area leads to a failure to discriminate the solution, as shown in
column 2.  Shrinking the temporal width of the pulses causes a loss
of the spectral selectivity needed to distinguish between the
excitonic and biexciton resonances.  In order to exhibit the effects
of the unintended dynamics and of the pulse-shaping remedy, the first
three column have not included decoherence. The effect of shaping on
quantum computation is shown the third column \cite{supplemental}: the
possibility to distinguish between constant and balanced within the
decoherence time is recovered.   The robustness of our scheme is tested with the inclusion
of decoherence in column 4.  The pulse shaping can in principle be
applied to more than two excitons. There is still the question of
whether the pulse shaping is ``scalable'' to a system of $n$ qubits,
as n increases. The advantage of a quantum computer over a classical
one is the saving of effort from $2^n$ steps to order of powers in $n$
\cite{chuang}. We need to avoid using an algorithm which treats the
$n$ excitons locked together as a strongly interacting system. The
method \cite{divince} is to process in steps, each involving a few
excitons. Then, the overhead of shaping is at most a few powers of $n$
\cite{tian}, i.e., it does not spoil the exponential saving.

In conclusion, we resolve the basic physics issues raised by the
optical control of two antiparallel-spin excitons in a quantum
dot. The application of this idea to the control of two spin-polarized
excitons in a single dot is well within present experimental
capabilities, and we hope that our explicit design will stimulate a
quick realization of an elementary quantum control. Our solution of
pulse-shaping is a special case of quantum control of interacting
spins and may be applied to other physical systems, such as trapped
ions.  Future work would involve excitons in different dots. Several
schemes already exist for scalable system where qubits stored in an
array of dots can be swapped around in a distributed quantum computer
\cite{imamoglu00,wang}. Cavity Quantum Electrodynamics effects can be
used for the interconnection between dots, and have, in principle, the
additional potential of controlling the spontaneous emission of the
excitons.

LJS thanks Drs. D. Steel, I. Chuang, D. Gammon, and Hailin Wang for
helpful discussions. CP acknowledges discussions with
Dr. J. Fernandez-Rossier and the support by the Swiss NSF.  
This work was supported in part by the NSF Grant No. DMR 9721444 and
in part by DARPA/ONR N0014-99-1-109.

\section*{appendix}

\begin{figure} 
\centerline{\psfig{file=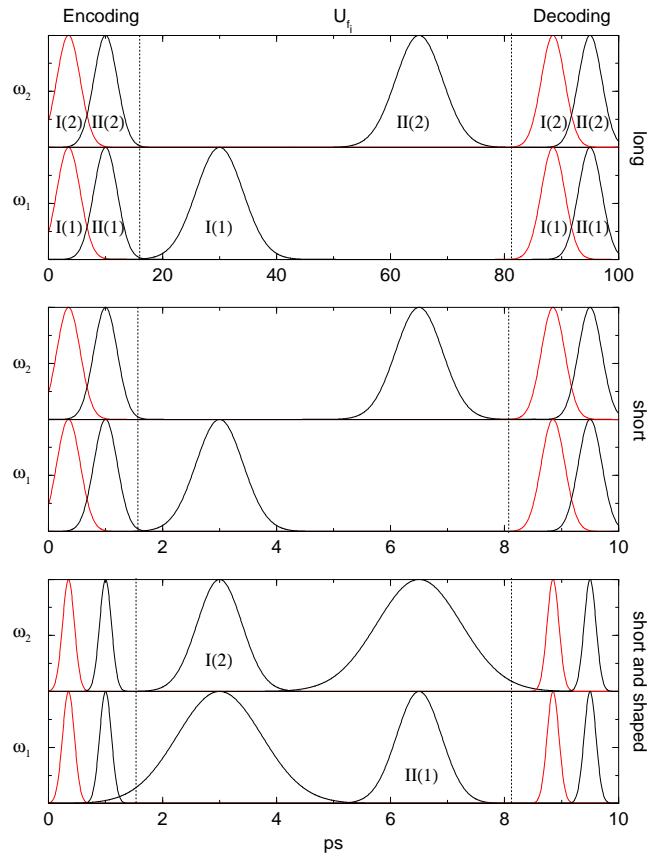,width= 8.6 cm,angle=
0}}\vspace{0.5 cm} 
\caption{Slowly varying envelopes of the pulse sequences used in the
quantum computation for the solution of the DJ problem in
Fig.\ref{computing}. The red (black) line indicates $\sigma +$
($\sigma -$) circular polarization. The two rows labeled by $\omega_i$
refers to the resonance frequency of the
pulses: $\omega_1=\epsilon_+=\epsilon_-$,$\omega_2=\epsilon_+-\Delta_{+-}=\epsilon_---\Delta_{+-}$.}
\label{supp1}
\end{figure}

\vspace{1.cm}
\begin{table}
\caption{Pulse sequences used in the computation, (1) and (2) indicate
the two components of a shaped pulse of the form ${\cal E}(t)$ $={\cal
E}_1(t)+{\cal E}_2(t)$ $={\cal E}_0(e^{-(t/t_1)^2-i\omega_1
t-i\phi_1}+e^{-(t/t_2)^2-i\omega_2 t + i \phi_2})~.$ The area $\theta$
is defined as $\hbar \theta=\int d {\cal E}(t)~dt$. $d{\cal E}_0 $ is
0.2 meV in the long pulses case and 2 meV for short and shaped pulses.
The pulse sequence of $U_{f_1}(U_{f_4})$ is identical to the
$U_{f_2}(U_{f_3})$ except for a change of sign in the phase $\phi_j$.}
\begin{tabular}{c|c|c|c|c|c|c}
&Pol & $\phi_j$ & $\omega_j$&Long pulses & Short pulses &Shaped pulses \\ 
& & & & ps ($\theta$) & ps ($\theta$) & ps($\theta$) \\
\hline \hline 
\multicolumn{7}{c}{ ENCODING }\\ 
\hline \hline 
I  (1)&$\sigma +$&$-\pi/2$& $\omega_1$ & 2.92 ($\pi$/2) &0.29 ($\pi$/2) & 0.15 (0.81)\\ 
I  (2)&$\sigma +$&$-\pi/2$& $\omega_2$ & 2.92 ($\pi$/2) &0.29 ($\pi$/2)  &0.15 (0.81)\\ 
II (1)&$\sigma -$& $\pi/2$ & $\omega_1$ & 2.92 ($\pi$/2) &0.29 ($\pi$/2) &0.15 (0.81)\\ 
II (2)&$\sigma -$& $\pi/2$ & $\omega_2$ & 2.92 ($\pi$/2) &0.29 ($\pi$/2) &0.15 (0.81)\\ 
\hline \hline
\multicolumn{7}{c}{ DECODING}\\ 
\hline \hline
I  (1)&$\sigma +$& $\pi/2$& $\omega_1$ & 2.92 ($\pi$/2) &0.29 ($\pi$/2) &0.15 (0.81)\\ 
I  (2)&$\sigma +$& $\pi/2$& $\omega_2$ & 2.92 ($\pi$/2) &0.29 ($\pi$/2) &0.15 (0.81)\\ 
II (1)&$\sigma -$& $\pi/2$ & $\omega_1$ & 2.92 ($\pi$/2) &0.29 ($\pi$/2) &0.15 (0.81)\\ 
II (2)&$\sigma -$& $\pi/2$ & $\omega_2$ & 2.92 ($\pi$/2) &0.29 ($\pi$/2) &0.15 (0.81)\\ 
\hline \hline
\multicolumn{7}{c}{$U_{f_2}$}\\
\hline 
I (1)& $\sigma -$ & $-\pi/2$ & $\omega_1$ & 5.83 ($\pi$) & 0.58 ($\pi$)  & 1.05 (5.68) \\ 
I (2)& $\sigma -$ & $ \pi/2$ & $\omega_2$ &  & & 0.56 (3.03)\\ 
II(2)& $\sigma -$ & $-\pi/2$ & $\omega_2$ & 5.83 ($\pi$) & 0.58 ($\pi$)& 1.05 (5.68)\\ 
II(1)& $\sigma -$ & $ \pi/2$ & $\omega_1$ & & & 0.56 (3.03)\\ 
\hline
\multicolumn{7}{c}{$U_{f_3}$}\\
\hline 
I (1)& $\sigma -$ & $-\pi/2$ & $\omega_1$ & 5.83 ($\pi$) & 0.58 ($\pi$)  & 1.05 (5.68) \\ 
I (2)& $\sigma -$ & $ \pi/2$ & $\omega_2$ & & & 0.56 (3.03)\\ 
II(2)& $\sigma -$ & $ \pi/2$ & $\omega_2$ & 5.83 ($\pi$) & 0.58 ($\pi$)& 1.05 (5.68)\\ 
II(1)& $\sigma -$ & $-\pi/2$ & $\omega_1$ & & & 0.56 (3.03)\\ 
\end{tabular}
\label{tab}
\end{table}



\begin{references}
\bibitem{supplemental} Details of pulse sequences are given in the Appendix.
\end{references}
\end{document}